\documentclass[12pt]{article}
\usepackage{latexsym}
\newcommand{\cB}{{\mathcal B}}
\newcommand{\cH}{{\mathcal H}}
\newcommand{\1}{{\mathbf 1}}
\newcommand{\rP}{{\rm PAIRS}}
\newcommand{\ra}{{\rm rank}}
\newcommand{\T}{{\rm Tr}}

\begin{document}

\title{ ``PARTIAL'' FIDELITIES}

\author{Armin Uhlmann }

\date{Institut f.~Theoretische Physik, Universit\"at Leipzig\\
Augustusplatz 10/11, D-04109 Leipzig}

\maketitle

\begin{abstract}
For pairs $\omega$, $\rho$,
of density operators on a finite dimensional Hilbert
space of dimension $d$ I call $k$-fidelity the $d - k$
smallest eigenvalues of $| \sqrt{\omega} \sqrt{\rho} |$.
$k$-fidelities are jointly concave in $\omega, \rho$. This
follows from representing them as infima over linear functions.
For $k=0$ known properties of fidelity and transition
probability are reproduced. Partial fidelities
characterize equivalence classes of pairs of density operators
which are partially ordered in a natural way.
\end{abstract}

\section{Introduction}
For two unit vectors, $\psi$ and $\varphi$, of an Hilbert space
the quantity $|\langle \psi, \varphi\rangle|^2$ is their transition
probability. It is the squared modulus of their transition amplitude,
$\langle \psi, \varphi\rangle$. Assume the state of the quantum
system is $|\psi\rangle\langle\psi|$.
A von Neumann measurement, designed
to decide whether the quantum system is in the state
$|\varphi\rangle\langle\varphi|$, prepares this state
with probability $|\langle \psi, \varphi\rangle|^2$.\\
Notice further that two pairs of unit vectors are unitarily
equivalent iff they enjoy equal transition probabilities.

All that becomes more complex, \cite{fidel}, if two density
operators, $\rho_1$ and $\rho_2$, are considered on a Hilbert space
$\cH$, and the quantum system is in state, say, $\rho_1$. The algebra
of operators on $\cH$ will be called $\cB$.
One can choose vectors $\psi_j$ in the direct product $\cH \otimes \cH$
such that
\begin{equation} \label{pf1}
\T \, A \rho_j = \langle \psi_j, (A \otimes \1) \psi_j \rangle,
\quad A \in \cB, \, j = 1, 2
\end{equation}
The transition probability between $\psi_1$ and $\psi_2$ is not
determined by the pair $\rho_1$, $\rho_2$. But running through all
possible arrangements (\ref{pf1}), the numbers $|\langle \psi_2,
\psi_1\rangle|^2$ fill completely an interval $[0, p]$ of  real
numbers. The largest one, the upper bound of this interval, is
called {\em transition probability between $\rho_1$ and $\rho_2$}
and is denoted by $P(\rho_1, \rho_2)$. Thus a von Neumann
measurement in $\cH \otimes \cH$ can cause a transition $\rho_1
\mapsto \rho_2$ with a probability bounded by $P(\rho_1, \rho_2)$.
The bound can be reached by suitable measurements in the
larger system.\\
Now I call attention to possibilities to
characterize $P$ intrinsically, i.~e. without leaving the quantum
system in question. The first one comes rather directly from
(\ref{pf1}). Let us call {\em transition functional from $\rho_1$
to $\rho_2$} every linear functional on $\cB$ of the form
\begin{equation} \label{pf2}
A \longrightarrow \T \, \nu A :=
\langle \psi_2, (A \otimes \1) \psi_1 \rangle
\end{equation}
which arises from a setting (\ref{pf1}). The operators $\nu$ may be
called {\em transition operators from $\rho_1$ to $\rho_2$}.
Generally, $\nu$ is not Hermitian: Exchanging the roles of
$\psi_1$ and $\psi_2$ the operator $\nu$ becomes its Hermitian
adjoint, $\nu^*$.\\
Now (\ref{pf2}) is a transition functional if and only if
\begin{equation} \label{pf3}
| \, \T \, A_1 \nu A_2^* \, |^2 \leq ( \T \, A_1 \rho_1 A_1^* ) \,
( \T \, A_2 \rho_2 A_2^* ), \quad A_i \in \cB
\end{equation}
and it follows from the definition of $P$ that
\begin{equation} \label{pf4}
P(\rho_2, \rho_1) = \max | \, \T \, \nu \, |^2
\end{equation}
where one takes the maximum over all transition operators from
$\rho_1$ to $\rho_2$.
Calculating the maximum in (\ref{pf4}) is a standard exercise
with a well known outcome. Before writing it down
I would like to explain the following:\\
The transition probability is separately concave in every one of
its arguments.
However, taking the root of $P$, the concavity properties become
dramatically enhanced: $\sqrt{P}$ is jointly concave
\cite{AlUh:84}.
In the following the square root of the transition probability
will be called {\em fidelity} and will be denoted by $F$,
essentially following a proposal
of Richard Jozsa\footnote{Jozsa introduced the word {\em fidelity} for
the transition probability. Its present usage is not unique.
I think
the peculiar properties of $\sqrt{P}$ need an extra notation
anyway.}. Thus
\begin{equation} \label{pf5}
F(\omega, \rho) := \sqrt{P(\omega, \rho)} = \T \bigl( \rho^{1/2}
\omega \rho^{1/2} \bigr)^{1/2}
\end{equation}
The assertion that $F$ is jointly concave is seen from
\begin{equation} \label{pf6}
F(\omega, \rho) = {1 \over 2} \inf
\bigl( \T \, A \omega + \T \, A^{-1} \rho \bigr),
\quad A > 0, \, \, A^{-1} \in \cB
\end{equation}
which is the finite dimensional version of a representation of
$\sqrt{P} = F$ as an infimum of linear functionals, valid for
pairs of states on von Neumann and on C$^*$-algebras, see
\cite{AU99}. The representation is related to another one of
equal generality estimating $P(\omega, \rho)$ from above by
the product of
Tr$\omega A$ and Tr$\rho A^{-1}$, with $A$ an invertible positive
operator, see \cite{inf1} for a partial result and \cite{inf2}
for the C$^*$-case in full generality. For finite dimensions
these well know results are reproduced by setting $k = 0$ in
the expressions (\ref{infk2}) and (\ref{infk3}) below.

As a matter of fact, the equality of $F$  (or of $P$) for two
pairs of density operators do not imply their unitary equivalence.
This pleasant feature, valid for pure states, is missing for
mixed ones.
Looking at (\ref{pf6}) one may wonder whether it is
not possible to get a whole series of concave invariants by taking
other suitable sets of operators than the invertible positive
operators in the expression (\ref{pf6}). To give an affirmative
answer belongs to the issues of the present paper. By
the partial fidelities one gets a reasonable classification of
pairs of density operators, coarser than unitary
equivalence would give.

All what follows remain in finite dimensions.
By modifying certain settings and by adding new arguments,
Peter M.~Alberti,
\cite{Alberti}, was able to extend essential parts of what follows
to von Neumann Algebras.
His results are particularly satisfying for type II$_1$.

\section{$k$--Fidelities}
Let $\cH$ be a finite-dimensional Hilbert space and $d = \dim \cH$.\\
The {\em spectrum}, spec$(A)$, of an operator $A$ is the
family of roots of the polynomial $\det(A - \lambda \1)$ counted
with their correct multiplicities. If the spectrum is real
we assume the set spec$(A)$ {\em decreasingly ordered}.
This convention applies to every diagonalizable operator
with real eigenvalues
and in particular to every Hermitian one. Consider now
\begin{equation} \label{spectrum}
\hbox{spec}\bigl((\sqrt{\omega} \rho \sqrt{\omega})^{1/2}\bigr)
=
\hbox{spec}\bigl((\sqrt{\rho} \omega \sqrt{\rho})^{1/2}\bigr)
=
\{ \, \lambda_1 \geq \lambda_2 \geq \dots, \geq \lambda_d \, \}
\end{equation}
so that, according to (\ref{pf5}),
the sum of the lambdas is the fidelity. The spectrum (\ref{spectrum})
is equal to the ordered singular numbers of
$\sqrt{\omega} \sqrt{\rho}$ and of $\sqrt{\rho} \sqrt{\omega}$.\\
I define  partial fidelities simply by summing
up parts of the spectrum (\ref{spectrum}).\\
{\bf Definition:} \, For $0 \leq k \leq d-1$
\begin{equation} \label{kfidel}
F_k(\omega, \rho) := \sum_{j > k} \lambda_j, \quad
k=0, 1, \dots, d-1 .
\end{equation}
If $k \geq d$  then $F_k = 0$. For the time being
$F_k$ will be called {\em $k$-th partial fidelity}, or simply
{\em $k$-fidelity} of the pair $\omega$ and $\rho$.\\
An important point is to add: I do not necessarily require
that $\rho$ and $\omega$ have trace one. Indeed, on a finite
dimensional Hilbert space (\ref{kfidel}) is naturally defined for
pairs from the cone of positive operators and
\begin{equation} \label{scale}
\sqrt{c} \, F_k(\omega, \rho)
= F_k(c \omega, \rho) = F_k(\omega, c \rho), \quad c > 0
\end{equation}
for positive real numbers $c$. Notice the properties:\\
a) $F_k$ is symmetric in its arguments.\\
b) $F_0$ is just the fidelity $F$.\\
c) For pairs of pure density operators it is $F_k = 0$ for $k > 0$.\\
d) If $F_k \neq 0$ then rank$(\omega) > k$ and
rank$(\rho) > k$ necessarily.\\
e) $F_k$ is unitarily invariant, i.e.~invariant by the
simultaneous transformation
$\rho \to U \rho U^*$ , $\omega \to U \omega U^*$.\\
However, a deeper justification for the definition above is in\\
{\bf Theorem 1}\\
{\em The partial fidelities are concave functions of the pairs }
$\{ \omega, \rho \}$ :
\begin{equation} \label{concave}
\sum_j p_j F_k(\omega_j, \rho_j) \leq F_k(\sum_j p_j \omega_j,
\sum_i p_i \rho_i)
\end{equation}
{\em for any probability vector $p_1, p_2, \dots$ and arbitrary
pairs $\{ \omega_i, \rho_i \}$.} $\Box$

\noindent
The theorem is a consequence of a new relation representing $F_k$
as an infimum of linear functionals quite similar to (\ref{pf6}).
It estimates partial fidelities linearly as close as possible
from above. To get the announced representation
I am going to define the set PAIRS which consists
of all pairs $\{A, B\}$ of positive Hermitian
operators, $A$, $B$, such that
\begin{equation} \label{pairs1}
 ABA = A, \quad BAB = B
\end{equation}
Let $\{A, B\}$ be such a pair. It follows $(AB)^2 = AB$ immediately.
Because $Q = AB$ is a product of two positive operators it is
diagonalizable. On the other hand we see $Q^2 = Q$ so that its
spectrum consists of zeros and ones. Therefore, the trace of $Q$
is equal to the rank of $Q$. Now (\ref{pairs1}) says $QA=A$ and
$BQ=B$ implying that the ranks of $A$ and $B$ cannot be larger
than the rank of $Q$. Now $Q=AB$ shows that neither the rank of
$A$ nor the rank of $B$ can be less than that of $Q$. Altogether
we have:\\
{\bf Lemma 1:} \, For all  $\{A, B\} \in \rP$
\begin{equation} \label{pairs2}
\ra(A) = \ra(B) = \ra(AB) = {\rm Tr} \, AB
\end{equation}
is an integer called {\em rank of the pair} $\{A, B\}$. $\Box$\\
{\bf Definition:} PAIRS$_m$ consists of all pairs from PAIRS of
rank $m$.\\ The promised representation of the $k$-th fidelities
is in the following theorem.\\

\noindent
 {\bf Theorem 2}\\
  {\em Let $m + k = \dim \cH$. Then}
\begin{equation} \label{infk2}
F_k(\omega, \rho) = {1 \over 2} \, \inf \Bigl( \,
\T \, A \omega + \T \, B \rho \Bigr), \quad
\{A, B \} \in \rP_m.
\end{equation}

\noindent
One can deduce from (\ref{infk2}) the following inequality:
\begin{equation} \label{infk3}
F_k(\omega, \rho)^2 = \inf
\Bigl(\T \, A \omega \Bigr)  \Bigl(\T \, B \rho \Bigr),
\quad \{A, B \} \in \rP_m.
\end{equation}
The point is that with $\{A, B\}$ also
$\{\lambda A, \lambda^{-1} B\}$ is contained in  PAIRS$_m$
for $\lambda >0$. After this trivial substitution, the right hand
side of (\ref{infk2}) is of the form $\lambda a + \lambda^{-1}b$.
Taking the infimum of over $\lambda$ results in $2\sqrt{ab}$,
and (\ref{infk3}) is derived from (\ref{infk2}). (\ref{infk3}),
suitably reformulated, is known on C$^*$-algebras if $k=0$,
see \cite{inf2}.

Theorem 1 is a consequence of theorem 2.
We shall prove theorem 2 in the next section, at first assuming
$\rho$ invertible, (which would be sufficient for theorem 1).
Then, by continuity arguments, we can allow for all $\omega$ and $\rho$.
However, before going into the proof,
we have to look at a "hidden" symmetry of the $k$-fidelities.

\section{The symmetry group of the $k$-fidelities}
Let us denote by $\Gamma$
the multiplicative group of all invertible operators acting on $\cH$.
With $X \in \Gamma$ we define the {\em $X$-transform} of a pair
$\{ \omega, \rho \}$ by
\begin{equation} \label{transf1}
\{ \omega, \rho \}^X := \{ X \omega X^*, (X^{-1})^* \rho X^{-1} \}
\end{equation}
The transformations create orbits of $\Gamma$ in the set
of pairs. Two pairs, $\{ \omega, \rho \}$
and $\{ \omega', \rho' \}$, are called {\em $\Gamma$-equivalent}
iff there is $X \in \Gamma$ such that
$\{ \omega', \rho' \}$ is the $X$-transform
of $\{ \omega, \rho \}$.\\
{\bf Lemma 2:} \,
The $k$-fidelities of $\Gamma$-equivalent pairs are equal
for every $k$
\begin{equation} \label{inv1}
F_k(\omega, \rho) = F_k(\omega', \rho') \, \hbox{ if } \,
\{ \omega', \rho' \} = \{ \omega, \rho \}^X
\end{equation}
For the proof   we start with the identity
$$
(X \omega X^*) (X^{-1})^* \rho X^{-1} = X \omega \rho X^{-1}
$$
saying that the spectrum of $\omega \rho$ is an invariant for
$\Gamma$-equivalent pairs. It suffices to show that
\begin{equation} \label{spectrum2}
\hbox{spec}(\omega \rho) = \hbox{spec}(\rho \omega)
= \{ \, \lambda_1^2, \lambda_2^2, \dots, \lambda_d^2 \, \}
\end{equation}
follows from (\ref{spectrum}). By
$\sqrt{\rho} (\omega \rho) (\sqrt{\rho})^{-1} =
\sqrt{\rho} \omega \sqrt{\rho}$ this is true for invertible $\rho$.
The assumption of invertibility can be removed
by continuity.$\Box$\\
%%%%%%%%%%%%%%%%%%%%%%%%%%%%%%%%%%%%%%
Substituting in (\ref{transf1}) $\rho = X^* \tau X$ we may rewrite
(\ref{inv1}) as
\begin{equation} \label{inv1a}
F_k(\omega, X^* \tau X) = F_k( X \omega X^*, \tau)
\end{equation}
In (\ref{inv1a}), by continuity, also $X$ need not be invertible.

%%%%%%%%%%%%%%%%%%%%%%%%%%%%%%%%%%%%%%
Only for the purpose of the the following proof
we abbreviate
the right-hand-side of (\ref{infk2}) by  $G_k(\omega, \rho)$,
$$
G_k(\omega, \rho) := {1 \over 2} \, \inf \Bigl( \,
\T \, A \omega + \T \, B \rho \Bigr), \quad
\{A, B \} \in \rP_m.
$$
We  observe that also $G_k$ allows for $\Gamma$-invariance.
Indeed, with a pair $\{A, B \}$ its transform
\begin{equation} \label{transf2}
\{A, B \}_X := \{ X^* A X, X^{-1} B (X^{-1})^* \}
\end{equation}
is also in PAIRS$_m$, and we get the trace identities
$$
\T \omega A = \T \omega' A', \quad \T \rho B  = \T \rho' B'
$$
whenever
$$
\{A', B' \} = \{A, B \}_X, \quad \{ \omega', \rho' \} =
\{ \omega, \rho \}^X
$$
Therefore,
$$
G_k(\omega, \rho) = G_k(\omega', \rho') \, \hbox{ if } \,
\{ \omega', \rho' \} = \{ \omega, \rho \}^X
\eqno(a)
$$
Now we start the {\em proof of theorem 2}. As we just have seen,
{\em both} sides of (\ref{infk2}) do not change along
a $\Gamma$-orbit of $X$-transforms (\ref{transf1}).

Step 1 in the proof is to show $F_k \leq G_k$.  To this end we first
assume a pair $\{ \omega, \rho \}$ with an
invertible $\rho$.  We transform the given pair according to
(\ref{transf1}) by $X = \sqrt{\rho}$.
The new pair is $\{ \omega', \1 \}$ with
$\omega' = \sqrt{\rho} \omega \sqrt{\rho}$.  By (a) and lemma 2
it suffices to prove $F_k \leq G_k$ for the new
pair, i.~e. we estimate $G_k(\omega', \1)$ from below.\\
We choose a pair $\{ A, B \}$ from
PAIRS$_m$ arbitrarily.  Let $\phi_1, \phi_2, \dots$ be an eigenbasis of
$A$ and $a_1, a_2, \dots$ the corresponding eigenvalues.  By
sandwiching $ABA=A$ between these eigenvectors of $A$ one gets
$$
\langle \phi_i, B \phi_k \rangle =
a_i^{-1} \delta_{ik} \, \, \hbox{for} \, \, i, k \geq m
$$
Now we can write
$$
\T \, A \omega' + \T
\, B = \sum_1^m a_i \langle \phi_i, \omega' \phi_i \rangle + \sum_1^m
a_i^{-1} + \sum_{j > m} \langle \phi_j, B \phi_j \rangle
$$
With
positive reals $a$ and $x$ it holds $ax + a^{-1} \geq 2 \sqrt{x}$.
Using that inequality to estimate the first two sums from below and
neglecting the last term, we arrive at
$$
\T \, A \omega' + \T \, B \geq
2 \sum_1^m \sqrt{ \langle \phi_i, \omega' \phi_i \rangle }
$$
The square root is concave.
Hence, see \cite{A+U}, equ. 1-46,
$$
\sum_1^m \sqrt{\langle \phi_i, \omega' \phi_i \rangle }
\geq \sum_1^m \langle \phi_i,
\sqrt{\omega'} \phi_i \rangle \geq F_k(\omega, \rho)
$$
The last
inequality sign is an estimation of the $m$ smallest eigenvalues (due to
Fan and Horn) and respecting  $F_k(\omega', \1) = F_k(\omega, \rho)$.
It results
\begin{equation} \label{below}
{1 \over 2} \, \inf \Bigl( \, \T \, A \omega + \T \, B \rho \Bigr) \geq
F_k(\omega, \rho)
\end{equation}
at first for pairs $\{\omega', \1 \}$
and then, by $\Gamma$-invariance, for all pairs $\{ \omega, \rho \}$
with invertible $\rho$.  However, both sides of (\ref{below}) are
continuous in $\omega$ and $\rho$.  Thus step one terminates in
the validity of
(\ref{below}) for all $\omega$ and all $\rho$. The inequality is
equivalent to $G_k \geq F_k$.\\
In step 2
we show $G_k \leq F_k$ at first for invertible $\rho$.  As above we
reduce the problem by $\Gamma$-invariance to that of a pair
consisting of $\omega'$ and $\1$.  We now choose
$\phi_1, \dots, \phi_m$ to be eigenvectors of $\omega'$ belonging
to the $m$ smallest eigenvectors
of $\omega'$.  The latter are $\lambda_{k+1}^2, \dots, \lambda_{k+m}^2$.
by lemma 2. Define
$$
A' = \sum_1^m a_i |\phi_i\rangle\langle\phi_i|, \quad
B' = \sum_1^m a_i^{-1} |\phi_i\rangle\langle\phi_i|
$$
and consider
$$
\T A' \omega' + \T B' =
\sum_{j=1}^m a_j \lambda_{j+k}^2 + \sum_1^m a_j^{-1}
$$
If $\lambda_{j+k} > 0$ we choose $a_j = \lambda_{j+k}^{-1}$.
Otherwise we set $a_j = c^{-1} > 0$ arbitrarily.  If $n$ of the $m$
eigenvalues $\lambda_{j+k}$ are zero, then our convention implies
$$
\T A' \omega' + \T B' = 2 \sum_{j=1}^m \lambda_{j+k} + nc =
2 F_k(\omega', \1) + nc
$$
and, hence, $G_k \leq F_k + nc$.
Since $c$ can be made arbitrarily small we arrive at the wanted
inequality $G_k \leq F_k$. Now, relying on $\Gamma$-invariance
(\ref{inv1}) and (a), the inequality is shown true
for all pairs of invertible density operators.\\
Combining step one and two we see:
$F_k(\omega, \rho) = G_k(\omega, \rho)$ if both arguments
are invertible.  Hence $F_k$ is concave for these pairs.
But $F_k$ is a continuous function of $\omega$ and $\rho$
by (\ref{kfidel}).  Therefore, $F_k$ is jointly concave
and theorem 1 is valid.\\
But one knows that a concave
function is semi-continuous from below, see \cite{Roc70}, theorem 10.2,
where semi-continuity from above is stated for convex functions.
Because $F_k$ is continuous and concave it dominates every function
which is concave and coincides for convexly inner points with $F_k$.
This means $F_k \geq G_k$ always.
Now step one of the proof provides $F_k = G_k$. $\Box$

\section{Equivalence and partial order}
It is tempting to collect pairs of positive (density) operators
into equivalence classes according to their partial fidelities.
For the purpose of the present paper we call two pairs
{\em equivalent}, and we write
\begin{equation} \label{sim}
\{ \omega, \rho \} \sim \{ \omega', \rho' \},
\end{equation}
iff their $k$-fidelities are equal,
$F_k(\omega, \rho) = F_k(\omega', \rho')$
for $k = 0, 1, \dots, d-1$. The relation $\sim$ is an
equivalence relation. Notice that
$\{ \omega, \rho \} \sim \{ \rho, \omega, \}$.
Generally, an equivalence class
contains a lot of $\Gamma$-orbits. But there is an important
exception:\\
{\bf Lemma 3:} \,
If both operators, $\omega$ and $\rho$, are invertible,
the equivalence class of $\{\omega, \rho\}$ consists exactly
of all pairs $\{\omega, \rho\}^X$, $X \in \Gamma$ .$\Box$

Proof: The assumption is valid if and only if 0 does not belong
to the eigenvalues (\ref{spectrum}). This takes place if the
smallest one is different from zero, hence iff $F_{d-1} \neq 0$.
Hence, if the assumption of the lemma is valid for one member
of an equivalence class, then it is true for all members.
Let $\{\omega_1, \rho_1\}$ be in the equivalence class
of $\{\omega, \rho\}$.
Transforming the latter by
$X = \sqrt{\rho}$ and the former by $X_1 = \sqrt{\rho_1}$
 by the receipt (\ref{transf1}) results in accordingly
$\Gamma$-equivalent pairs
$\{\omega', \1 \}$ and $\{\omega'_1, \1 \}$. Being in the same
equivalence class, $\omega'$ and $\omega'_1$
have to have equal eigenvalues and they are even  unitarily
equivalent. Thus all the pairs considered belong to the same
$\Gamma$-orbit.  $\Box$

Let us write
$\{\omega', \rho' \} \leq \{\omega, \rho \}$ if both,
$\omega - \omega'$ and $\rho - \rho'$, are positive
operators. A simple example
is as follows: Write $\omega = \omega' + \omega_0$,
$\rho = \rho' + \rho_0$, and assume orthogonality between
$\omega'$ and $\rho_0$ and between $\rho'$ and $\omega_0$, i.~e.
$\omega_0 \rho' = 0$, $\rho_0 \omega' = 0$. Then
$\{ \omega, \rho \}$ and $\{ \omega', \rho' \}$
belong to the same equivalence class.
To see what we can learn from
$\{\omega', \rho' \} \leq \{\omega, \rho \}$
generally, we proceed in two steps,
$\{\omega', \rho' \} \leq \{\omega, \rho' \}$ and
$\{\omega, \rho' \} \leq \{\omega, \rho \}$. Consider the
 second one. It implies
$\sqrt{\omega} \rho'\sqrt{\omega} \leq \sqrt{\omega} \rho \sqrt{\omega}$
and, because taking the square root does not destroy the
inequality,
$$
\bigl(\sqrt{\omega} \rho'\sqrt{\omega} \bigr)^{1/2}
\leq
\bigl(\sqrt{\omega} \rho \sqrt{\omega} \bigr)^{1/2}
$$
The sums of its $m$ smallest eigenvalues, which are the partial
fidelities, obey the same inequality. Further, if the traces
of both positive operators happen to be equal, the operators
themselves have to be equal one to another. In repeating
the arguments for the first step and combining both, we arrive
at\\
{\bf Lemma 4:} \, If
\begin{equation} \label{L41}
\{ \omega, \rho \} \geq \{ \omega', \rho' \} \,
\end{equation}
then
\begin{equation} \label{pgeq}
F_k(\omega, \rho) \geq F_k(\omega', \rho'),
\quad k = 0, 1, \dots, d-1
\end{equation}
If in addition to (\ref{L41}) $F(\omega, \rho) = F(\omega', \rho')$
is true, then all partial fidelities must be equal in pairs, and the
two pairs belong to the same equivalence class:
$\{ \omega, \rho \} \sim \{ \omega', \rho' \}$. $\Box$

Given $\omega$, $\rho$,
Alberti \cite{mpairs} has shown, even in the C$^*$-category,
that there is one and only one pair $\{ \omega_0, \rho_0 \}$
which enjoys the same transition probability, (and, therefore,
the same fidelity), and which is minimal with
respect to $\geq$. This {\em minimal pair} satisfies
$$
\{ \omega_0, \rho_0 \} \leq \{ \omega', \rho' \}
$$
whenever
$$
\{ \omega', \rho' \} \leq \{ \omega, \rho \} \,\, \hbox{ and } \,\,
F(\omega', \rho' ) = F(\omega, \rho )
$$
is valid.\\
We see that every equivalence class contains a minimal
pair and, therefore, a $\Gamma$-orbit of minimal pairs. It is
tempting to believe that there is only one minimal $\Gamma$-orbit
in every equivalence class of pairs. But I do not know whether
that conjecture is true.\\

Now one may go a step further, anticipating the ideas of
majorization, \cite{M+O}, or those of partially ordering orbits
belonging to certain classes of transformations, \cite{A+U}.
To do so, let us  call
 $\{ \omega_1, \rho_1 \}$ {\em F--dominated} by
 $\{ \omega_2, \rho_2 \}$  iff
\begin{equation} \label{F-dom}
F_k(\omega_1, \rho_1) \leq F_k(\omega_2, \rho_2), \quad
k = 0, 1, 2, \dots
\end{equation}
From theorem 1 we get the following\\
{\bf Corollary:} \,
If $\{ \omega_2, \rho_2 \}$ is contained in the convex hull of
the $\sim$equivalence class of $\{ \omega_1, \rho_1 \}$ then
(\ref{F-dom}) takes place. $\Box$

We thus get a new partial ordering (or majorization tool) for pairs
of positive (density) operators which seems worthwhile to
investigate.
There is a link, indeed a morphism, to singular number majorization.
Denote by sing$(B)$ the decreasingly ordered singular numbers of
the operator $B$, that is
$$
 \hbox{sing}(B) = \hbox{spec}(\sqrt{B^*B}) =
\hbox{spec}(\sqrt{BB^*}) =  \hbox{sing}(B^*),
$$
and by sing$[B]$ the set of all operators $C$ with
sing$(C)$ = sing$(B)$, the {\em singular number class} of $B$.
In particular
$$
\hbox{spec}((\sqrt{\omega} \rho \sqrt{\omega})^{1/2}) =
\hbox{sing}(\sqrt{\rho} \sqrt{\omega})
$$
There are many useful rules governing the partial order of the
singular number classes, see \cite{M+O}, 9.E and \cite{A+U},
2.4 (theorem 2-8), 2.5. With them one easily proves\\
{\bf Lemma 5:} \, The following items are mutually equivalent:\\
a)  $\{ \omega_1, \rho_1 \}$ is F--dominated by
 $\{ \omega_2, \rho_2 \}$.\\
b) $\sqrt{\omega_2} \sqrt{\rho_2}$ is contained in the convex hull
of sing$[\sqrt{\omega_1} \sqrt{\rho_1}]$.\\
c) There are finitely many operators $A_i$, $B_i$,  all with operator
norms not exceeding 1, such that
$$
\sqrt{\omega_2} \sqrt{\rho_2} = \sum A_i \sqrt{\omega_1}
\sqrt{\rho_1} B_i
$$

\section{More about PAIRS}
It is our aim to get some insight into the structure of PAIRS.
Let $\{ A, B\} \in$ PAIRS$_m$ with $0 < m \leq d = \dim \cH$.
$k$ is defined by $k+m=d$.
Let us write $A$, $B$ as block matrices with respect to an
eigenvector basis of $A$ as in the proof of theorem 2.
Then, with is a positive $m \times m$ matrix $A_{11}$,
\begin{equation} \label{block1}
A = \pmatrix{A_{11} & 0 \cr 0 & 0 \cr}, \quad
B = \pmatrix{B_{11} & B_{12} \cr B_{21} & B_{22} \cr}
\end{equation}
Here $B_{11}$ is $m \times m$,  $B_{12}$ is $m \times k$,
and $B_{22}$ is $k \times k$.
 The equation $ABA=A$ results in
$B_{11} = A_{11}^{-1}$. Having this in mind, one gets from $BAB=B$
\begin{equation} \label{block2}
B_{11} = A_{11}^{-1}, \quad B_{22} = B_{21} A_{11} B_{12}, \quad
B_{12}^* = B_{21}
\end{equation}
Notice that $B_{12}$ can be chosen arbitrarily: Given the first
member, $A$, of the pair, $B$ depends freely on $km$ complex
parameters.

There is a further representation of the pairs in PAIRS$_m$.
Call $2m$ vectors, $\psi_1, \psi_2, \dots, \psi_m$,
$\varphi_1, \varphi_2, \dots, \varphi_m$, a
{\em bi-orthogonal system of length $m$} if
$$
 \langle \psi_i, \varphi_j \rangle = \delta_{ij}, \quad
i, j = 1, 2, \dots, m
\eqno(A)
$$
Together with $m$ positive numbers, $a_1, a_2, \dots, a_m$,
we obtain from (A) a pair of operators
$$
A = \sum a_i |\psi_i \rangle\langle \psi_i|, \quad
B = \sum a_i^{-1} |\varphi_i \rangle\langle \varphi_i|
\eqno(B)
$$
for which $ABA=A$ and $BAB=B$ can be
checked. Let us prove that {\em every pair from} PAIRS$_m$
{\em can be gained by this procedure.}

Let $\{ A, B \} \in$ PAIRS$_m$. Because $AB$ is diagonalizable
with eigenvalues 0 and 1, there is
$X \in \Gamma$ such that $X AB X^{-1}$ is a Hermitian projection
operator. Hence the operators
$$
A_1 = X A X^*, \quad B_1 = (X^{-1})^* B X^{-1}
$$
commute. Therefore there is a
representation
$$
A_1 = \sum a_i |\phi_i \rangle\langle \phi_i|, \quad
B_1 = \sum a_i^{-1} |\phi_i \rangle\langle \phi_i|
$$
with $m$ orthonormal vectors $\phi_1, \dots, \phi_m$. But
$$
\psi_i = X^{-1} \phi_i, \quad \varphi_i = X^* \phi_i
$$
is bi-orthogonal with length $m$. Transforming $A_1$
and $B_1$ back to $A$ and $B$ gives the desired representation
of the pair.\\
The bi-orthonormal system (A) of (B) can be chosen {\em balanced} :
$$
\langle \psi_i, \psi_i\rangle = \langle \varphi_i, \varphi_i\rangle,
\quad i = 1, \dots, m
\eqno(C)
$$
Indeed, the necessary changes in the norms can be compensated
by adjusting the $a_i$. Now we insert (B) into the right hand side of
(\ref{infk2}) and observe
$$
\T \, A \omega + \T \, B \rho =
\sum_1^m \bigl(  a_j < \psi_j | \omega | \psi_j > +
 a_j^{-1} < \varphi_j | \varrho | \varphi_j > \bigr)
$$
By varying the free parameters $a_j$ we arrive at\\
{\bf Theorem 3}\\
 {\em Let $m + k = \dim \cH$. Then}
\begin{equation} \label{infk4}
F_k(\omega, \rho) = \inf \sum_{i=1}^m
\sqrt{\langle \psi_i, \omega \psi_i\rangle
\langle \varphi_i, \rho \varphi_i\rangle}
\end{equation}
{\em where the infimum runs through all balanced bi-orthogonal
systems of length} $m$.  $\Box$

Finally, assume the infimum in (\ref{infk2}) is attained by
$\{A, B\} \in$ PAIRS$_m$,
\begin{equation} \label{minimum1}
F_k(\omega, \rho) = {1 \over 2}
\Bigl( \, \T \, A \omega + \T \, B \rho \Bigr),
\quad m + k = \dim \cH
\end{equation}
If we vary the minimizing pair, the first variation must vanish,
$$
\Bigl({d \over ds} \Bigr)_{s=0} c(s), \quad c(s) =
(\T \, A_s \omega + \T \, B_s \rho)
$$
where, with $X_s = \exp s Y$ and any operator $Y$,
$$
A_s = X_s^* A X_s, \quad B_s = X_s^{-1} B (X_s^*)^{-1}
$$
We perform the first derivative and obtain
$$
\Bigl({d \over ds} \Bigr)_{s=0} A_s = Y^* A + A Y,
\quad
\Bigl({d \over ds} \Bigr)_{s=0} B_s = - Y B - B Y^*
$$
After inserting in $\dot c(0) = 0$ and an rearrangement it results
$$
{\rm Tr} \, Y (A \rho - \omega B)^* +
{\rm Tr} \, Y^* (A \rho - \omega B) = 0
$$
As $Y$ could be chosen arbitrarily, we arrive at
\begin{equation} \label{minimum2}
 A \rho = \omega B
\end{equation}
as a necessary condition for the validity of (\ref{minimum1}).

Is there $\{A, B\} \in$ PAIRS$_m$ fulfilling (\ref{minimum1})
and minimizing (\ref{infk2})? If we can $\Gamma$-transform
$\omega, \rho$, to the form $\tau, \tau$, we are
done. Indeed, we then can choose a projection operator $P_m$
onto the $m$ smallest eigenvalues of $\tau$ and we get
$$
F_k(\tau, \tau) = {\rm Tr} \, P_m \tau, \quad
\{P_m, P_m \} \in \rP_m
$$
i.~e. the problem is solved in that case. Now, if $\omega$
and $\rho$ are both invertible, there is a unique positive $X$
such that
\begin{equation} \label{minimum3}
X \omega X = X^{-1} \rho  X^{-1} := \tau, \quad X > 0
\end{equation}
The choice (\ref{minimum3}) ensures (\ref{minimum1}) with
$A = XP_mX$, $B = X^{-1} P_m X^{-1}$. To get $X$ one has to
solve $X^2 \omega X^2 = \rho$. There is
a unique positive solution $X$ which is the square root
of the geometric mean \cite{PW75} between $\rho$ and $\omega^{-1}$.
$$
X^2 = \omega^{-1/2} \bigl( \omega^{1/2} \rho \omega^{1/2} \bigr)^{1/2}
\omega^{-1/2}
$$
as one can convince oneself by inserting into $X^2 \omega X^2 = \rho$.

\section*{Acknowledgement}
I like to thank P.~M.~Alberti for several good advices and
 B.~Crell, C.~Fuchs, H.~Narnhofer, and W.~Thirring for
stimulating discussions.
I am thankful  for support to the Erwin Schr\"odinger Institute for
Mathematical Physics, Vienna, and, during the Workshop on Complexity,
Computation and the physics of information,
to the Isaac Newton Institute for
Mathematical Sciences, Cambridge, and to the European Science
Foundation.

\end{document}